\documentclass[conference]{IEEEtran}
\IEEEoverridecommandlockouts

\usepackage{cite}
\usepackage{amsmath,amssymb,amsfonts}

\usepackage{graphicx}
\usepackage{textcomp}
\usepackage{xcolor}
\def\BibTeX{{\rm B\kern-.05em{\sc i\kern-.025em b}\kern-.08em
    T\kern-.1667em\lower.7ex\hbox{E}\kern-.125emX}}

\usepackage{amsthm}

\usepackage{amsmath}

\usepackage{graphicx}
\usepackage{tikz}
\usetikzlibrary{arrows,fit,shapes}
\usetikzlibrary{calc}
\tikzset{fontscale/.style = {font=\relsize{#1}}}
\usepackage{pgfplots}
\usepackage{algpseudocode}
\usepackage{algorithm}
\usepackage{caption}
\usepackage{multirow}
\usepackage{subcaption}
\usepackage{relsize}
\usepackage{url}
\usepackage{adjustbox}

\usepackage{pgfplots}
\usepackage{filecontents}
\pgfplotsset{compat=newest}

\usepackage{cite}

\usepackage{proof}
\usepackage{capt-of}
\usepackage{mathtools}
\usepackage{amsmath,amsfonts,amssymb}
\usepackage{microtype}
\usepackage{hyperref}
\usetikzlibrary{plotmarks}

\makeatletter
\tikzset{
    scale plot marks/.is choice,
    scale plot marks/false/.code={
        \def\pgfuseplotmark##1{\pgftransformresetnontranslations\csname pgf@plot@mark@##1\endcsname}
    },
    scale plot marks/true/.style={},
    scale plot marks/.default=true
}

\makeatother

\newcommand{\task}[1]{\tau_{#1}}

\newcommand{\vertex}[2]{v_{#1#2}}

\newcommand{\vertexset}[1]{\mathcal{V}_{#1}}

\newcommand{\edgeset}[1]{\mathcal{E}_{#1}}

\newcommand{\srr}{\mathsf{s}}
\newcommand{\fin}{\mathsf{f}}

\newtheorem{example}{Example}

\newcommand{\he}[1]{[Houssam : \textcolor{red}{#1}]}

\author{Houssam-Eddine Zahaf$^*$, Nicola Capodeici$^+$ \\   $^*$Université de Nantes, LS2N, France \\ $^+$Unimore university, HiPeRT Lab, Italy}

\begin{document}

\title{Building  Time-Triggered Schedules for typed-DAG Tasks with alternative implementations}

\maketitle

\thispagestyle{plain}
\pagestyle{plain}

\begin{abstract}
  \label{sec:abstract}
  Hard real-time systems like image processing, autonomous driving,
  etc.  require an increasing need of computational power that
  classical multi-core platforms can not provide, to fulfill with
  their timing constraints. Heterogeneous Instruction Set Architecture
  (ISA) platforms allow accelerating real-time workloads on
  application-specific cores (e.g. GPU, DSP, ASICs) etc. and are
  suitable for these applications. In addition, these platforms
  provide larger design choices as a given functionnality can be
  implemented onto several types of compute elements.

  HPC-DAG (Heterogeneous Parallel Directed Acyclic Graph) task model
  has been recently proposed to capture real-time workload execution
  on heterogeneous platforms. It expresses the ISA heterogeneity, and
  some specific characteristics of hardware accelerators, as the
  absence of preemption or costly preemption, alternative
  implementations and on-line conditional execution.

  In this paper, we propose a time-table scheduling approach to
  allocate and schedule a set of HPC-DAG tasks onto a set of
  heterogeneous cores, by the mean Integer Linear Programming
  (ILP). Our design allows to handle heterogeniety of resources,
  on-line execution costs, and a \emph{faster} solving time, by
  exploring gradually the design space.
\end{abstract}


\begin{IEEEkeywords}
Real-time partitioning, heterogeneous architecture, unrelated, preemption, time-table.
\end{IEEEkeywords}

\section{Introduction}
\label{sec:introduction}

Cyber-physical embedded systems are increasingly complex and demand
more and more powerful computational hardware
platforms. COTS\footnote{Commercial Off The Shielf} vendors provide
hardware platforms featuring multi-core CPU hosts with a number
hardware accelerators, in order to support timing constraints of
complex real-time applications with machine learning and image
processing software modules.

Typically, these platforms feature different ISA architectures, and
different characteristics for the different cores. For example, the
Jetson AGX platform features 8 CPU with ARM V8, Volta GPU (Graphical
Processing Unit), a Deep Learning Accelerators (DLA), a Programmable
Vision Accelerator (PVA). The differences between the different types
of cores include not only ISA, but as well as the task scheduling,
i.e. CPU can be usually scheduled with preemptive scheduling
approaches where preemption costs can be negligible, while recent
GPUs enable preemptive scheduling but with very variable
costs\footnote{The cost varies from some micro-seconds to couple of
  milliseconds.}, while when a workloads starts executing on DLA or
PVA, the latter can not be preempted.

\he{Some bla bla about how good it is to express alternative, and
  conditionals}.
\he{Say that it is complex to make non-time consuming ILPs to solve
  the conditional problem, therefore they are not considered in this
  paper}

When it comes to scheduling, there are two major approaches for
designing and scheduling HPC-DAGs onto heterogeneous cores :
time-triggered and event-based approaches. In event-triggered systems,
tasks are scheduled as a consequence of events i.e. task activation,
deadlines, priority, etc.  therefore the schedule is not necessarily
known before run-time \cite{tc2021zahaf}. In contrast time-triggered
systems, the system designer precompute/or totally compute in offline
static execution schedule, i.e. time windows, for every job of every
task during the whole system lifetime. Time-triggered and event-based
approaches are incomparable as each exhibit advantages and
disadvantages with respect to each other.

In contrast to our previous work in which we focus on event-based
schedulers, this paper presents a time-triggered scheduler for a set
of HPC-DAG tasks onto a heterogeneous multicore platform.

\paragraph*{Contributions.}

\begin{enumerate}
\item Formulation of the problem using several ILP
\item Accelerating the analysis
\item taking into account heterogeneity
\item Adding partitioning
\item Adding global scheduling
\item taking into account preemption costs
\end{enumerate}


\section{System model}
\label{sec:system-model}

\subsection{Architecture model}
\label{sec:architecture-model}

A heterogeneous architecture is modeled as a set of \emph{execution
  engines} $\mathtt{Arch} = \{ e_{1}, e_2, \ldots, e_m\}$.  An
execution engine is characterized by 1) its execution capabilities,
(e.g. its Instruction Set Architecture), specified by the engine's
\emph{tag}, and 2) its scheduling policy. An engine's tag
$\mathtt{tag}(e_i)$ indicates the ability of a processor to execute
dedicated tasks.

As an example, a Xavier based platform such as the \emph{NVIDIA
  pegasus}, can be modeled using $16$ engines for a total
of five different engine tags: $8$ \textsf{CPU}s, $2$ \textsf{dGPU}s,
$2$ \textsf{iGPU}s, $2$ \textsf{DLA}s and $2$ \textsf{PVA}s.

Tags express the heterogeneity of modern processor architecture: an
engine tagged by \textsf{dGPU} (discrete GPU) or \textsf{iGPU}
(integrated GPU) is designed to efficiently run generic GPU kernels,
whereas engines with \textsf{DLA} tags are designed to run \emph{deep
  learning inference} tasks. A deep learning task can be implemented
and compiled to run on any engine, including CPUs and GPUs, however
its execution behavior will be different, and its execution time will
probably be lower when running on DLAs.

Therefore, we allow the designer to compile the same task on different
alternative engines with different trade-offs in terms of performance
and resource utilization, so to widen the space of possible
solutions. As we will see in the next section, the HPC-DAG model
supports \emph{alternative} implementations of the same code.

In this paper, we consider time triggered scheduling. For every engine
we build a schedule table, that dictates the scheduling at run-time.

\subsection{The HPC-DAG task model}
\label{sec:task-model}

A \emph{task} is a Directed Acyclic Graph (DAG), characterized by a
tuple $\tau = \{\mathsf{T}, \mathsf{D}, \mathcal{N}, \edgeset{}\}$,
where: $\mathsf{T}$ is the period (exact inter-arrival time between two
consecutive activation of task $\tau$); $\mathsf{D}$ is the relative
deadline; The set of all the nodes is denoted by
$\mathcal{N}$

An edge $e(n_i, n_j) \in \edgeset{}$ models a precedence constraint
(and related communication) between node $n_i$ and node $n_j$.  The
set of \emph{immediate predecessors} of a node $n_j$, denoted by
$\mathsf{pred}(n_j)$, is the set of all nodes $n_i$ such that there
exists an edge $(n_i, n_j)$. The set of \emph{predecessors} of a node
$n_j$ is the set of all nodes for which there exist a path toward
$n_j$.  If a node has no predecessor, it is a \emph{source node} of
the graph. In our model we allow a graph to have several source nodes.
In the same way we define the set of \emph{immediate successors} of
node $n_j$, denoted by $\mathsf{succ}(n_j)$, as the set of all nodes
$n_k$ such that there exists an edge $(n_j, n_k)$, and the set of
\emph{successors} of $n_j$ as the set of nodes for which there is a
path from $n_j$. If a node has no successors, it is a \emph{sink node}
of the graph, and we allow a graph to have several sink nodes.

A node $n_i$ can be a sub-task or an alternative node.  A sub-task
$v \in \vertexset{}$ is the basic computation unit.  It represents a
block of code to be executed by one of the engines of the
architecture. A sub-task is characterized by:
\begin{itemize}
\item A tag $\mathtt{tag}(v)$ which represents the engines where it is
  eligible to execute. A sub-task can only be allocated onto an engine
  with the same tag;
\item A worst-case execution time $C(v)$ when executing the sub-task
  on the corresponding engine processor.
\item $\mathsf{P}(v)$ Maximum number of preemptions allowed for
  sub-task $v$.
\item $\mathsf{}{cost}(v)$ : The cost to split (preempt) task $v$
\end{itemize}

An alternative node $a \in \mathcal{A}$ represents alternative
implementations of parts of the graph/task. During the configuration
phase, our methodology selects one between many possible alternative
implementations of the program by selecting only one of the outgoing
edges of $a$ and removing (part of) the paths starting from the other
edges for a given job. This can be useful when modeling sub-tasks than
can be executed on different engines with different execution
costs. The selected edge may differ from an activation to another
according the system state when the job is executed, in contrast to
our previous work \cite{tc2021zahaf} where the same alternative
configuration is selected during the system lifetime.


An alternative nodes always has at least 2 outgoing edges, so they
cannot be sinks.

\begin{example}
\label{example:task-model-1}
Consider the DAG task described in 
Figure~\ref{fig:tau_spec_2}. Each sub-task node is labeled by the
sub-task id and engine tag.  An Alternative node are denoted by square
boxes. The black boxes denotes corresponding junction nodes for
alternatives.

\begin{figure}[h]
  \centering
  \begin{subfigure}{.5\columnwidth}
    \centering
    \resizebox{\textwidth}{!}{\begin{tikzpicture}[rotate = 90]
  \def\d{1.7}
  \node at (0,0) [circle,draw] (v_1) {\tiny $v_{1}^{\mathsf{CPU}}$};
  \node at (\d,0) [circle, draw] (v_2) {\tiny $v_{2}^{\mathsf{CPU}}$};
  \node at (\d/2,-1) [rectangle,draw] (A_1) {\textcolor{black}{A}};
  \node at (\d/2,-3*\d) [rectangle,draw,fill] (A_C) {\textcolor{black}{\scriptsize F}};
  \node at (-0.1,-\d) [circle, draw, dashed] (v_3) {\textcolor{white}{\tiny $v_{3}^{\mathsf{dGPU}}$}};
    \node at (1.5,-\d-0.1) [circle, draw] (v_3p) {\tiny $v_{3}^{\mathsf{dGPU}}$};
  \node at (\d,-1.7*\d) [circle, draw] (v_4) {\tiny $v_{4}^{\mathsf{DLA}}$};
  \node at (\d,-2.4*\d) [circle, draw] (v_5) {\tiny $v_{5}^{\mathsf{dGPU}}$};
  \node at (\d/2,-3.7*\d) [circle, draw] (v_8) {\tiny $v_{8}^{\mathsf{CPU}}$};
  \node at (-0.5,-1.75*\d) [circle, draw] (v_6) {\tiny $v_{6}^{\mathsf{DLA}}$};
  
  \node at (0.5,-1.75*\d) [circle, draw] (v_7) {\tiny $v_{7}^{\mathsf{dGPU}}$};
  \draw [<-](A_1) -- (v_1);
  \draw [<-](A_1) -- (v_2);
  \draw [->](A_1) -- (v_3p);
  \draw [->](v_3p) -- (v_4);
  \draw [->](A_1) -- (v_3);
  \draw [->](A_1) -- (v_3);
  \draw [->](v_4) -- (v_5);
  \draw [->](v_5) -- (A_C);
  \draw [->](v_6) -- (A_C);
  \draw [->](v_3) -- (v_7);
  \draw [->](v_3) -- (v_6);
  \draw [->](v_7) -- (A_C);
  \draw [->](A_C) -- (v_8);
\end{tikzpicture}}
    \caption{}\label{fig:tau_spec_2}
  \end{subfigure}
  \quad\quad        
  \begin{subfigure}{.4\columnwidth}
    \centering
    \resizebox{\textwidth}{!}{\begin{tikzpicture}[rotate = 90]
  \def\d{1.7}
  \node at (0,0) [circle,draw] (v_1) {\tiny $v_{1}^{\mathsf{CPU}}$};
  \node at (\d,0) [circle, draw] (v_2) {\tiny $v_{2}^{\mathsf{CPU}}$};
  \node at (\d/2,-0.5*\d) [circle,draw,dashed] (F_2) {\textcolor{white}{\tiny $v_{2}^{\mathsf{CPU}}$}};
  \node at (\d/2,-2.5*\d) [circle, draw] (v_8) {\tiny $v_{8}^{\mathsf{CPU}}$};
  
  \node at (0,-1*\d) [circle, draw] (v_6) {\tiny $v_{6}^{\mathsf{DLA}}$};
  \node at (\d,-1*\d) [circle, draw] (v_7) {\tiny $v_{7}^{\mathsf{dGPU}}$};
  \draw [<-](F_2) -- (v_1);
  \draw [<-](F_2) -- (v_2);
  \draw [->](F_2) -- (v_7);
  \draw [->](F_2) -- (v_6);
   \draw [->](v_6) -- (v_8);
   \draw [->](v_7) -- (v_8);
 \end{tikzpicture}

}
    \caption{}\label{fig:concrete_1}
  \end{subfigure}
  \caption{Task specification and concrete tasks}
\end{figure}
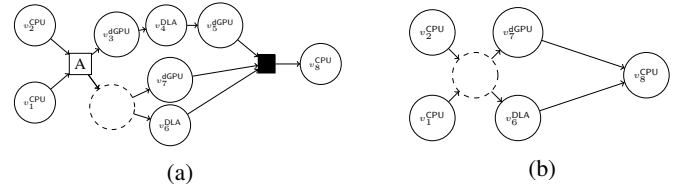


Sub-tasks $\vertex{}{1}^{\mathsf{CPU}}$ and
$\vertex{}{2}^{\mathsf{CPU}}$ are the sources (entry points) of the
DAG. $\vertex{}{1}^{\mathsf{CPU}}, \vertex{}{2}^{\mathsf{CPU}}$ are
marked by the \textsf{CPU} tag and can run concurrently: during the
off-line analysis they may be allocated to the same or to different
engines in a consecutive or parallel time windows. Sub-task
$\vertex{}{4}^{\mathsf{DLA}}$ has an outgoing edge to
$\vertex{}{5}^{\mathsf{dGPU}}$, thus sub-task
$\vertex{}{5}^{\mathsf{dCPU}}$ cannot start its execution before
sub-task $\vertex{}{4}^{\mathsf{DLA}}$ has completed. Sub-tasks
$\vertex{}{1}^{\mathsf{CPU}}$ and $\vertex{}{2}^{\mathsf{CPU}}$ have
each one outgoing edge to the alternative node $A$. Thus, $\task{}$
can continue the execution either:
\begin{enumerate}
\item by following $\vertex{}{3}^{\mathsf{dGPU}}$ and then
  $\vertex{}{4}^{\mathsf{DLA}}$,$\vertex{}{5}^{\mathsf{dGPU}}$ and
  finishing its instance on $\vertex{}{8}^{\mathsf{CPU}}$;
\item or by following a dummy node (denoted as dashed) to execute in
  parallel $\vertex{}{6}^{\mathsf{DLA}}$ or
  $\vertex{}{7}^{\mathsf{dGPU}}$, and finishing its instance on
  $\vertex{}{8}^{\mathsf{CPU}}$.
\end{enumerate}

The two patterns are alternative ways to execute the same
functionalities at different costs. 
Figure~\ref{fig:concrete_1} represents one of the possible executions
of task $\tau$, where during the analysis, alternative execution
pattern $(\vertex{}{3}^{dGPU},\vertex{}{4}^{DLA},\vertex{}{5}^{dGPU})$
has been dropped.
\end{example}

We consider a \emph{periodic task} model, therefore parameter
$\mathsf{T}$ represents the exact inter-arrival time between two
instances of the same concrete task. When an instance of a task is
activated at time $t$, all source sub-tasks are simultaneously
activated. All subsequent sub-tasks are activated upon completion of
their predecessors, and sink sub-tasks must all complete no later than
time $t+\mathsf{D}$. We assume \emph{constrained deadline tasks}, that
is $\mathsf{D} \leq \mathsf{T}$.

\section{Time table construction using ILP}
\label{sec:analysis}

In this section, we describe how the time table for the HPC-DAG can be
constructed. We consider both global and partitioned approaches. In
the partitioned approach a sub-task is allocated to a given core
during the system lifetime.  Therefore, all execution time-windows of
every job of a given sub-task are reserved on the same core, while in
the global approach, different time intervals of the same job might be
executed on different cores.

An optimal solution requires to assign every time unit from 0 to the
hyper-period, on every core to a job or to \emph{idle}.  Therefore, the
design space to explore is large, and finding an optimal solution is
an extremely time consuming operation. In this work, we explore the
design space gradually. Rather than assigning processor time to jobs,
we assign execution intervals to core.

Our approach selects a number of intervals per sub-task and increases
it at each iteration.  This allows to explore each iteration it
explores a subset of the design space instead of exploring the whole
design space at once, as described in Algorithm \ref{alg:full}.

Our algorithm computes first the maximal possible number of intervals
($\mathsf{nb\_it}$) per job as $\mathsf{nb\_it}=2^{it}$, where $it$
denotes the iteration number starting from 0. Further, it iteratively
increments $it$ by $1$. At each iteration, our algorithm builds an ILP
and invokes the CPLEX solver.

If the ILP solver founds a feasible solution (i.e. a feasible
schedule) our algorithm stops on \textsf{success}, otherwise the
system increases the number of intervals, and iteratively build and
solve a new ILP with new variable list. When the maximum number if
intervals is reached for every sub-task, and no feasible solution is
found, the system aborts on \textsf{fail}. In the rest of this
section, we describe how every ILP is built.

\he{I need to think about if the algorithm is optimal or not? (doubt not)}

\begin{algorithm}[t]
  \caption{Time-Table construction using ILPs} \label{alg:full}
  \begin{algorithmic}[1]
    \State {\bf Input:} Taskset $\mathcal{T}$, Architecture $\mathcal{A}$, Method : {\textsf{GLOBAL} or \textsf{PARTITIONED}}
    \State {\bf Output:} Time Table requirement task  $\tau^*$
    \State it=0
    \While {$(\mathbf{not}~\mathsf{stop}~\mathbf{and}~\mathbf{not}~\mathsf{feasible})$}
             \State ${\sf nb\_int = 2^{it}}$
             \State {\sf build\_variables(nb\_it);}
             \State {\sf build\_objective();}
             \State {\sf build\_sufficiency\_constraints();}
             \State {\sf build\_per\_proc\_overlapping\_constraints();}
             \State {\sf build\_per\_job\_overlapping\_constraints();}
             \State {\sf build\_finish\_before\_migrate\_constraints();}
             \State {\sf build\_precedance\_constraints();}
             \State {\sf build\_non\_migration\_constraints();}
             \If {({\sf METHOD ==  PARTITIONED})}
                   \State {\sf build\_partitioning\_constraints();}
             \EndIf
             \State {\sf feasible =  solve}();
             \If {$\sf (feasible  )$}
                  \State {\sf save\_time\_table }
                  \State \textbf{return } {\sf SUCCESS}; 
             \EndIf
             \State it+=1
             \State \textsf{update\_stop\_condition(stop)}
      \EndWhile
      \State \textbf{return}  {\sf FAIL}
  \end{algorithmic}
\end{algorithm}

\subsection{Building variables}
\label{sec:building-variables}

At each iteration $it$ is incremented and the number of intervals per
sub-task changes. Hence, the first step to build our ILP is to define
the variables list.

Our algorithm selects the number of intervals per sub-task at
iteration $it$ as
$\mathsf{nb\_it}(v) = \max\{\mathsf{nb\_it}, \mathsf{P}(v)
\}$. Therefore, sub-tasks that are executing onto a non-preemptive
engine, $\mathsf{nb\_it}(v)$ is set to $1$. For the sub-tasks
executing on in a fully preemptive engine, the number of intervals is
not defined in prior, $\mathsf{nb\_it}(v)$ is set to the maximum
between $\mathsf{nb\_it}$ and the sub-task worst case execution
time. In the first iteration, every job must execute into a exactly
one interval (i.e. non-preemptively), even those that are fully
preemptive as $it=0$.

For every interval, two variables are defined : $\srr$ and $\fin$ to
denote starting and finishing times respectively.

For every task, for every sub-task, for every job, for every interval,
for every core, we generate $\fin$ and $\srr$ variables. Therefore,
every variable is characterized by 5 indexes such that

$\srr_{i,j,k,l,m}$ (respectively $\fin_{i,j,k,l,m}$) denotes the
starting time (respectively finishing time) of the $l^{th}$ interval
of the $k^{th}$ job of sub-task $v_{i,j}$ which is executing on core
$m$. These variables and the relative constraints are generated only
for cores where a given sub-task is allowed to execute, i.e. a
sub-task that is meant to run on a GPU have no intervals defined CPU
cores.

Variables $\srr_{i,j,k,l,m}$ and $\fin_{i,j,k,l,m}$ are bounded by the
task activation, and task deadlines.

Therefore:

\begin{equation}
  \label{eq:var_bounds}
  k\cdot \mathsf{T}_i \leq  \srr_{i,j,k,l,m}  \leq  \fin_{i,j,k,l,m} \leq k\cdot \mathsf{T}_i + \mathsf{D}_i 
\end{equation}

When a job execution is split to several intervals, its execution time
is inflated by a penalty cost.

The procedure $\mathsf{build\_variables}$ (Line 6) allows to build the
variables and updates the task execution time as follows:

\[
  C(v)' = C(v) + \mathsf{nb\_it}(v) * \mathsf{cost}(v)
\]

\subsection{Objective function}
\label{sec:objective-function}

Once the list of variables are defined, our algorithm proceeds to
build the objective function to maximize the slack for each
job
. Therefore, the objective function is set to maximize the differences
between finishing and starting time for every interval, as follows :

\begin{equation}
  \mathsf{Maximize} \sum_{i=1}^n\sum_{j=1}^{|\mathcal{V}_i|}\sum_{k=1}^{H/T_i} \sum_{l=1}^{\mathsf{nb\_it}(v_{i,j})} \sum_{m=1}^{M^{(\mathsf{tag}(v_{i,j}))}} \fin_{i,j,k,l,m} - \srr_{i,j,k,l,m}
\end{equation}

Where :

\begin{itemize}
\item $n$ : is the number of task in the system  
\item  $|\mathcal{V}_i|$ The number of sub-tasks of task $\tau_i$
\item ${H/T_i} $ The number of jobs between 0 and H the hyper period
  for task $\tau_i$
\item ${\mathsf{nb\_it}(v_{i,j})}$ The number of intervals in the
  current iteration, computed in the previous step.
\item ${M^{(\mathsf{tag}(v_{i,j}))}}$ The indexes of cores having the same
  tag as the current job.
\end{itemize}

\subsection{Precedance constraints}
\label{sec:prec-constr}

To enforce the respect of precedance constraints, it is sufficient to
express that the successors of a any sub-task, can start before it
ends. In other words, any sub-task starting time must be greater to
all its immediate predecessors finishing time. For sub-task $v_{i,j}$,
multiple constraints can be generated as follows:

\begin{align}
  \forall & k \in \{0, \cdots, \frac{H}{\mathsf{T}_i} \},  & \forall &l ~ \in   \mathsf{nb\_it}(v_{i,j}),  & \forall &m \in M^{tag(v_{i,j})}  \notag  \\
     \forall &v_{i,j'} \in \mathsf{succ}(v_{i,j}), & \forall  & l' \in \mathsf{nb\_it}(v_{i,j'}), & \forall & m' \in M^{tag(v_{i,j'})} \notag  \\
  &&& ~~~~~~~~~\fin_{i,j,k,l,m} &\leq&~~~ \srr_{i,j',k,l',m'}
\end{align}

This constraints must be generated for every job, for every sub-task
in every task. Please notice that these constraints are generated
between successors within the same task instance (same $k$).

The precedance constraints between the different instances are
enforced by the variables $\mathsf{s}$ and $\mathsf{f}$ bounds
generated in variables building step.

\subsection{Interval sufficiency constraints}
\label{sec:interv-suff-constr}

After enforcing precedance constraints, we need to ensure that every
job of every sub-task will receive sufficiently processor time to
execute for its worst case execution time, including the costs of
splitting the task (preemption).  Therefore, the sum of interval
lengths for every instance, must be larger than the job worst case
execution time, as defined in following equation:

\[
\forall i, \forall j,  \forall k ,  \sum_l \sum_m  \fin_{i,j,k,l,m} - \srr_{i,j,k,l,m} \geq \mathsf{C}(v_{i,j})'
\]

\subsection{Overlapping constraints}
\label{sec:overl-constr}

At this stage, only the constraints to ensure a single task execution
have been defined. However, we need to ensure that the different tasks
executing on the same cores have a correct behavior.  Therefore, it is
required that the the same processor is not allocated to different
jobs at the same time and that the same job does not have reserved
intervals on different cores at the same time. We call these
constraints overlapping constraints.

First, we show how an o overlapping constraint can be
expressed for intervals $[\mathsf{s}, \mathsf{f}]$ and $[\mathsf{s}',\mathsf{f}']$ ,i.e :

\[
  [\srr     , \fin] \cap [\srr', \fin'] = \emptyset
\]

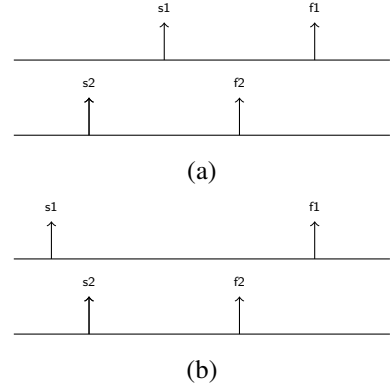
\begin{figure}[h]
  \centering
  \begin{tikzpicture}

\draw (0,0) -- (5,0);

\draw (0,1) -- (5,1);

\draw[->] (1,0) -- (1,0.5);

\def\x{3}
\def\y{0}
\draw[->] (\x,0+\y) -- (\x,0.5+\y);
\node at (\x,0.7+\y){\tiny $\mathsf{f2}$};

\def\x{2}
\def\y{1}
\draw[->] (\x,0+\y) -- (\x,0.5+\y);
\node at (\x,0.7+\y){\tiny $\mathsf{s1}$};

\def\x{4}
\def\y{1}
\draw[->] (\x,0+\y) -- (\x,0.5+\y);
\node at (\x,0.7+\y){\tiny $\mathsf{f1}$};

\def\x{1}
\def\y{0}
\draw[->] (\x,0+\y) -- (\x,0.5+\y);
\node at (\x,0.7+\y){\tiny $\mathsf{s2}$};
\node at (2.5,-0.5) {(a)};
\end{tikzpicture}
\begin{tikzpicture}

\draw (0,0) -- (5,0);

\draw (0,1) -- (5,1);

\draw[->] (1,0) -- (1,0.5);

\def\x{3}
\def\y{0}
\draw[->] (\x,0+\y) -- (\x,0.5+\y);
\node at (\x,0.7+\y){\tiny $\mathsf{f2}$};

\def\x{0.5}
\def\y{1}
\draw[->] (\x,0+\y) -- (\x,0.5+\y);
\node at (\x,0.7+\y){\tiny $\mathsf{s1}$};

\def\x{4}
\def\y{1}
\draw[->] (\x,0+\y) -- (\x,0.5+\y);
\node at (\x,0.7+\y){\tiny $\mathsf{f1}$};

\def\x{1}
\def\y{0}
\draw[->] (\x,0+\y) -- (\x,0.5+\y);
\node at (\x,0.7+\y){\tiny $\mathsf{s2}$};
\node at (2.5,-0.5) {(b)};
\end{tikzpicture}

  \caption{Example of overlapping }
  \label{fig:example_overlap}
\end{figure}

In Figure \ref{fig:example_overlap}, we presents two scenarios
scenarios where two interval $[\mathsf{s} , \mathsf{f}]$ and
$[\mathsf{s}', \mathsf{f}']$ overlap. Overlapping can be partial
i.e. only a part of the interval is shared as in Sub-figure (a) or a
total inclusion Sub-Figure (b).

\begin{figure}[h]
  \centering
  \begin{tikzpicture}

\draw (0,0) -- (5,0);

\draw (0,1) -- (5,1);

\def\x{2.5}
\def\y{0}
\draw[->] (\x,0+\y) -- (\x,0.5+\y);
\node at (\x,0.7+\y){\tiny $\mathsf{f2}$};

\def\x{3}
\def\y{1}
\draw[->] (\x,0+\y) -- (\x,0.5+\y);
\node at (\x,0.7+\y){\tiny $\mathsf{s1}$};

\def\x{4}
\def\y{1}
\draw[->] (\x,0+\y) -- (\x,0.5+\y);
\node at (\x,0.7+\y){\tiny $\mathsf{f1}$};

\def\x{1}
\def\y{0}
\draw[->] (\x,0+\y) -- (\x,0.5+\y);
\node at (\x,0.7+\y){\tiny $\mathsf{s2}$};
\node at (2.5,-0.5) {(a)};
\end{tikzpicture}
\begin{tikzpicture}

\draw (0,0) -- (5,0);

\draw (0,1) -- (5,1);

\def\x{4.5}
\def\y{0}
\draw[->] (\x,0+\y) -- (\x,0.5+\y);
\node at (\x,0.7+\y){\tiny $\mathsf{f2}$};

\def\x{0.5}
\def\y{1}
\draw[->] (\x,0+\y) -- (\x,0.5+\y);
\node at (\x,0.7+\y){\tiny $\mathsf{s1}$};

\def\x{2.5}
\def\y{1}
\draw[->] (\x,0+\y) -- (\x,0.5+\y);
\node at (\x,0.7+\y){\tiny $\mathsf{f1}$};

\def\x{3}
\def\y{0}
\draw[->] (\x,0+\y) -- (\x,0.5+\y);
\node at (\x,0.7+\y){\tiny $\mathsf{s2}$};
\node at (2.5,-0.5) {(b)};
\end{tikzpicture}

  \caption{Example of valid intervals}
  \label{fig:example_valid_overlap}
\end{figure}
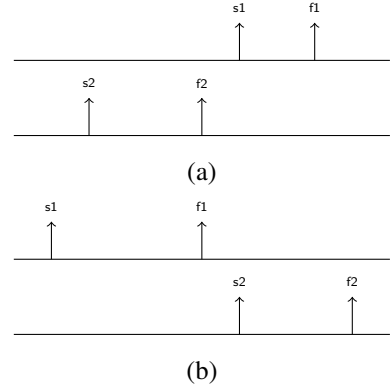

In Figure \ref{fig:example_valid_overlap}, we present the only two
cases to check that an overlapping is prohibited. Indeed, The first is
interval should either finish before the first one \textbf{or} start
after the second one as it is defined in Equation
(\ref{eq:constraint_nonlinear}).

\begin{align}
  \label{eq:constraint_nonlinear}
  \mathsf{f} \leq \mathsf{s'}~\mathbf{or}~\mathsf{f'} \leq \mathsf{s}  
\end{align}

Equation (\ref{eq:constraint_nonlinear}) ensures that two intervals do
not overlap, but need to be linearized.

We start first by linearizing the inequality
$\mathsf{f1} \leq \mathsf{s2}$ as follows :

\begin{equation}
  \label{eq:linear_ineq}
  \left\{
    \begin{array}{ll}
      \mathsf{f} - \mathcal{M} x_1 - \mathsf{s'} \leq 0  \\
      \mathsf{f} + \mathsf{M} - \mathcal{M} x_1 - \mathsf{s'} \geq 0 
    \end{array}
  \right.
\end{equation}

Where $\mathsf{M}$ is a very big number and $x_1$ is a binary variable
that the solver sets to $1$ if $\mathsf{f} \leq \mathsf{s'}$.

Similarly,  $\mathsf{f'} \leq \mathsf{s}$ is linearized as follows: 

\begin{equation}
  \label{eq:linear_ineq2}
  \left\{
    \begin{array}{ll}
      \mathsf{f'} - \mathcal{M} x_1 - \mathsf{s} \leq 0  \\
      \mathsf{f'} + \mathsf{M} - \mathcal{M} x_1 - \mathsf{s} \geq 0 
    \end{array}
  \right.
\end{equation}

As an interval can not be on the left and on the right at the same
time, we enforce that only one situation is selected by :

\begin{equation}
  \label{eq:singlecote}
  x_1 + x_2 = 1
\end{equation}

Now that we presented how two intervals can be enforced to not
overlap. We describe now how overlapping constraints are generated.
As described earlier, two types of overlapping must be prohibited :
(i) those on the same core, and (ii) those of the same job.

For the overlapping intervals of the same core, the constraints in
Equation (\ref{eq:linear_ineq}), (\ref{eq:linear_ineq2}),
(\ref{eq:singlecote}) must be generated of every couple of intervals allocated
to the same core having the same index of processor (index $m$ in the
variables name) as in the following equation :

\begin{align}
  \forall k, & \notag \\
  & &\forall i,\forall i',\forall j,  \forall j', \forall m, \forall m', \forall l,
    \forall l'  \\
  && [\srr_{i,j,k,l,m}, \fin_{i,j,k,l,m}] \cap&
[\srr_{i,j,k,l,m}, \fin_{i,j,k,l',m'}] = \emptyset \notag
\end{align}

For the overlapping intervals of the same job, the constraints in
Equation (\ref{eq:linear_ineq}), (\ref{eq:linear_ineq2}),
(\ref{eq:singlecote}) must be generated for every couple of intervals
of the job, i.e. having the same task, sub-task and job index
(i.e. the same $i, j,k$ in the variables name) as in the following equation :

\begin{align}
  \forall i, \forall j, \forall k, \forall m', \forall m, \forall l,
  \forall l', [\srr_{i,j,k,l,m}, \fin_{i,j,k,l,m}] \cap
  [\srr_{i,j,k,l,m}, \fin_{i,j,k,l',m'}] = \emptyset 
\end{align}

The core overlapping intervals constraints must be generated only when
considering the global approach. Indeed, in the partitioned approach
the same task will be always allocated on the same core, therefore the
job overlapping constraints are sufficient as tasks are forced to be
allocated to the same core, therefore can not have non-empty intervals
in the more than one core as it will be shown in the next section.

\subsection{Partitioning constraints}
\label{sec:part-constr}

At design time, one may allocate a sub-task to a particular core. To
enforce sub-task allocation without modifying the design of the ILP,
we ensure that only the intervals of the sub-task on the allocation
core can have positive lengths, while the length of all intervals on
the other cores have a length equal to zero.

Therefore, for all intervals of a given sub-task $v_{i,j}$ on a core $k$, we
define an allocation variable $a_{i,j,m}$. $a_{i,j,m}$ is equal to
$1$, if the interval length is strictly greater than zero, i.e:

\begin{equation}
  \label{eq:linear_non_linear2}
 a_{i,j,m} =  \left\{
    \begin{array}{ll}
      1, & \text{if} \exists k,l,m~\text{such } \fin_{i,j,k,l,m} > \srr_{i,j,k,l,m} \\
      0 & \text{Otherwise}
    \end{array}
  \right.
\end{equation}

Equation \eqref{eq:linear_non_linear2} is linearized as follows :

\begin{align}
  \fin_{i,j,k,l,m} - M \cdot a_{i,j,m}  - \srr_{i,j,k,l,m} \leq 0 \\
  \fin_{i,j,k,l,m} +(1-a_{i,j,m}) M     - \srr_{i,j,k,l,m} \geq 0
\end{align}

Please notice that the same variable $a_{i,j,m}$ is defined for all
intervals of the same sub-task and on the same core, therefore the
same variable $a_{i,j,m}$ is used for all intervals having the
processor index (i.e. $k$).

Further, we enforce that only a single variable is set equal to one :
\[
  \forall i, j, m\sum a_{i,j,l,m}  = 1; 
\]

\subsection{Handling alternative nodes }
\label{sec:hand-altern-nodes}

\section{Accelerating resolution}
\label{sec:accel-resol}

\bibliographystyle{IEEEtran}
\bibliography{mybib}

\end{document}